# New Possibilities for Observational Distinction Between Geometrical and Field Gravity Theories

© Yu. Baryshev[1,2]

[1] Astronomical Institute of the St.-Petersburg State University, St.-Petersburg, Russia
[2] Email: yuba@astro.spbu.ru

**Abstract:** Crucial observational tests of gravity physics are reviewed. Such tests are able to clarify the key question on the nature of gravitational interaction: is gravity the curvature of space? or is gravity a matter field in Minkowski flat space as other physical forces? Up to now all actually performed experiments do not allow to distinguish between these two alternatives in gravity physics. The existence of well-defined positive energy-momentum of the gravity field in Poincare-Feynman approach leads to radical changes in gravity physics and cosmology which may be tested by laboratory experiments and astrophysical observations. New possibilities for observational distinction between geometrical general relativity and field gravity theories are discussed. Among them: the contribution of the scalar repulsive force into Newtonian gravitational interaction, post-Newtonian translational motion of rotating bodies, gravitational deflection of light by small mass bodies, scalar gravitational radiation from spherically pulsating stars, existence of limiting radius, surface, magnetic field for massive bodies and absence of singularities and horizons for relativistic compact objects.

## 1. What is the nature of the gravitational interaction?

The central problem of the gravity physics is to understand the nature of the gravitational interaction. According to general relativity the gravity is a property of the geometry of the curved space, while in the frame of the field gravity theory the gravitational interaction is analogous to other physical forces. In the literature there is a statement that geometrical and field approaches are the same stories expressed by different languages. However as we demonstrated in the preceding paper (Baryshev 2008a) there are testable predictions which can distinct between these two alternatives in gravity physics.

Geometrical approach of the classical general relativity predicts such specific objects as singularities, black holes, and expanding space of Friedmann cosmological models. While in the field gravity theory there is no horizons and singularities, no expanding space, but there is the energy-momentum of the gravity field.

Weak gravity experiment performed by Nesvizhevsky et al. (2002; 2005) using freely falling ultra-cold neutrons, showed that the gravity force acts similarly to the usual electric force producing quantum energy levels for the micro-particles motion in the gravity field (Westphal at al. 2006). This experiment points to the field nature of the gravity force and lead us to look for other possibilities of testing the gravity physics.

## 2. Basic equations of the field approach

As we discussed in the preceding paper (Baryshev 2008a), within the framework of the field gravity theory the field equations have the form of the wave equation with the energy-momentum tensor of matter $T^{ik}$ as a fixed source of gravitational tensor potential $\psi^{ik}$ :

$$\left( \Delta - \frac{1}{c^2} \frac{\partial^2}{\partial t^2} \right) \psi^{ik} = \frac{8\pi G}{c^2} \left[ T^{ik} - \frac{1}{2} \eta^{ik} T \right] , \qquad (1)$$

where $T^{ik}$ contains all kinds of interacting matter including gravity, symmetric tensor potentials satisfy the Hilbert-Lorentz gauge conditions $\psi^{ik}_{,k} = 1/2 \ \psi^{,i}$ , and the scalar part of the field is $\psi = \eta_{ik}\psi^{ik}$.

The equation of motion of a test particle in a given fixed gravitational field $\psi^{ik}$ was derived by Baryshev (1986) in the form

$$A^i_k \frac{d(mcu^k)}{ds} = -mc B^i_{kl} u^k u^l , \qquad (2)$$

where $mcu^k = p^k$ is the 4-momentum of the particle,

$$A^i_k = \left(1 - \frac{1}{c^2}\psi_{ln}u^l u^n\right)\eta^i_k - \frac{2}{c^2}\psi_{kn}u^n u^i + \frac{2}{c^2}\psi^i_k \quad , \text{and} \quad B^i_{kl} = \frac{2}{c^2}\psi^i_{k,l} - \frac{1}{c^2}\psi^{,i}_{kl} - \frac{1}{c^2}\psi_{kl,n}u^n u^i \quad .$$

Here we present some solutions of these equations and calculate several observable effects which may be used as a crucial tests of the field gravity approach.

**3. Post-Newtonian predictions of the field gravity theory**

The field equations (1) and equations of motion (2) lead to important observable consequences of the field gravity theory. We consider some simple cases that demonstrate how to calculate weak-field predictions within FG. For solution of the field equations we use the method of iteration, where the non-linearity is taken into account by the iteration procedure.

*Weak gravity field of static spherically symmetric mass*

*Zero approximation – Newtonian limit.* For a spherically symmetric static weak field of a body with rest mass density $\rho_0(r)$ and total mass $M$, the zero approximation of the total EMT equals that of the matter

$$T^{ik}_{(m)} = diag(\rho_0 c^2, 0, 0, 0) \tag{3}$$

and the field equations have the usual Poisson's form

$$\Delta \psi^{ik} = 4\pi G \rho_0 \, diag(1, 1, 1, 1) \quad , \tag{4}$$

Solution of the field equations (4) is the Birkhoff's potential

$$\psi^{ik} = \varphi_N \, diag(1, 1, 1, 1) \quad , \tag{5}$$

where $\varphi_N = -GM/r$ is the Newtonian potential outside the gravitating body. We note again that $\psi^{ik}$ is a true tensor quantity in Minkowski space, hence the rules for contravariant and covariant components are usual.

The Birkhoff gravitational potential (5) can be expressed as the sum of the pure tensor and scalar components

$$\psi^{ik} = \frac{3}{2}\varphi_N \, diag(1, \frac{1}{3}, \frac{1}{3}, \frac{1}{3}) - \frac{1}{2}\varphi_N \, diag(1, -1, -1, -1) \quad , \tag{6}$$

Note that the scalar part of the Birkhoff potential is

$$\psi = \eta_{ik}\psi^{ik} = -2\varphi_N \quad , \tag{7}$$

which has opposite sign relative to Newtonian potential.

*First approximation – post-Newtonian limit.* In the first (post-Newtonian) approximation in accordance with the expression of the action integral ($S = S_{(m)} + S_{(int)} + S_{(g)}$) the total EMT of the system is equal to the sum of the three parts -- EMT for the matter, interaction and gravity field (Kalman 1961; Thirring 1961; Baryshev 1988):

$$T^{ik} = T^{ik}_{(p/m)} + T^{ik}_{(int)} + T^{ik}_{(g)} \quad , \tag{9}$$

Taking into account the Birkhoff potential (5) and using the expressions for the interaction EMT in the form

$$T^{ik}_{(int)} = \rho_0 \, \varphi_N \, diag(1, 1/3, 1/3, 1/3) \quad , \tag{10}$$

and the EMT of the gravity field in the form

$$T^{ik}_{(g)} = \frac{1}{8\pi G}(\vec{\nabla}\varphi_N)^2 \, diag(1, 1/3, 1/3, 1/3) \quad , \tag{11}$$

we find the total energy density ($T^{00}$) for the system gas + gravity in the form

$$T^{00} = T^{00}_{(p/m)} + T^{00}_{(int)} + T^{00}_{(g)} = (\rho_0 c^2 + e) + \rho_0 \varphi_N + \frac{1}{8\pi G}(\vec{\nabla}\varphi_N)^2 \ . \tag{12}$$

Here $(\rho_0 c^2 + e)$ gives the rest mass and kinetic (or thermal) energy densities, $\rho_0 \varphi_N$ is the negative interaction energy density, and $\frac{1}{8\pi G}(\vec{\nabla}\varphi_N)^2$ is the positive and localizable energy density of the gravitational field.

*Physical sense of the potential energy.* The total energy of the system in PN approximation will be

$$E_{(\Sigma)} = \int T^{00} dV = E_0 + E_k + E_p \ , \tag{13}$$

where $E_0 = \int \rho_0 c^2 \, dV$ is the rest-mass energy, $E_k = \int e \, dV$ is the kinetic energy, and $E_p$ is the classical potential energy, that equals the sum of the interaction and gravitational field energies:

$$E_p = E_{(int)} + E_{(g)} = \int (\rho_0 \varphi_N + \frac{1}{8\pi G}(\vec{\nabla}\varphi_N)^2) dV = \frac{1}{2}\int \rho_0 \varphi_N \, dV \ . \tag{14}$$

*The PN correction due to the energy of gravity field.* In the field approach a gravitating body is surrounded by a material gravitational field $\psi^{ik}$ whose mass-energy density is given by the 00-component of the EMT of the gravity field (11). In the PN approximation this leads to a nonlinear correction for the gravitational potential.

Outside the body the positive energy density of the gravitational field (the last term in eq.12) should be considered as the source in the field equation of the second order, then we get a nonlinear addition to Birkhoff's $\psi^{00}$ component

$$\psi^{00} = \varphi_N + \frac{1}{2}\frac{(\varphi_N)^2}{c^2} \ . \tag{15}$$

Corrections to other components do not influence the motion of particles in this approximation.

Very important that the positive energy density of the gravitational field is a measurable physical quantity within the framework of the field gravity theory as the additional non-liner term in equation (15).

### PN equations of motion and Poincare gravity force

*Poincare force.* In the post-Newtonian approximation we keep terms down to the order of $v^2/c^2 \propto \varphi_N/c^2 \ll 1$ in equation of motion of a test particle (2). For the PN accuracy we need calculations of the $\psi^{00}$ component with the order (15). Under these assumptions, from (2) for $i = \alpha$ we get an expression for the PN 3-dimensional gravity force (which we shall call the Poincare gravity force remembering his pioneer work in 1905 on the relativistic gravity force in flat space-time):

$$\vec{F}_{Poincare} = \frac{d\vec{p}}{dt} = -m_0\{(1 + \frac{3}{2}\frac{v^2}{c^2} + 3\frac{\phi}{c^2})\vec{\nabla}\phi - 3\frac{\vec{v}}{c}(\frac{\vec{v}}{c}\bullet\vec{\nabla}\phi) - 3\frac{\vec{v}}{c}\frac{\partial\phi}{c\partial t} + 2\frac{\partial\vec{\Psi}}{c\partial t} - 2(\frac{\vec{v}}{c}\times rot\, \vec{\Psi})\} \ , \tag{16}$$

where $\phi = \psi^{00}$, and $\vec{\Psi} = \psi^{0\alpha}$.

Taking into account the expression (15) for the 00-component of the gravitational potential, we get the corresponding PN 3-acceleration of a test particle:

$$\frac{d\vec{v}}{dt} = -(1 + \frac{v^2}{c^2} + 4\frac{\varphi_N}{c^2})\vec{\nabla}\varphi_N + 4\frac{\vec{v}}{c}(\frac{\vec{v}}{c}\bullet\vec{\nabla}\varphi_N) + 3\frac{\vec{v}}{c}\frac{\partial\varphi_N}{c\partial t} - 2\frac{\partial\vec{\Psi}}{c\partial t} + 2(\frac{\vec{v}}{c}\times rot\, \vec{\Psi}) \ , \tag{17}$$

*The work of the Poincare force.* From the $(i=0)$ component of the equation of motion (2) it follows an expression for the work of the Poincare force in PN approximation:

$$\frac{dE_k}{dt} = \vec{v}\bullet\vec{F}_{Poincare} = -m_0\vec{v}\bullet\{(1 - \frac{3}{2}\frac{v^2}{c^2} + 3\frac{\phi}{c^2})\vec{\nabla}\phi - 3\frac{\vec{v}}{c}\frac{\partial\phi}{c\partial t} + 2\frac{\partial\vec{\Psi}}{c\partial t}\} \ , \tag{18}$$

According to (18) the gravity force produce a work by changing kinetic energy of a test particle.

*Bi-component structure of the Newtonian force.* Inserting Birkhoff potential (6) in the equation of motion (2) we directly get that spin 2 part corresponds to attraction and spin 0 part gives the repulsion force. Indeed, in the Newtonian approximation we neglect all terms of order $v^2/c^2$ in the equation of motion (2), which gives for spatial components $i = \alpha$ the expression for the gravity force in the form, which demonstrate the contribution from each part of the Birkhoff potential:

$$\vec{F}_N = \frac{d\vec{p}}{dt} = -m\vec{\nabla}\varphi_N = -\frac{3}{2}m\vec{\nabla}\varphi_N + \frac{1}{2}m\vec{\nabla}\varphi_N = \vec{F}_{\{2\}} + \vec{F}_{\{0\}} = \frac{3}{2}\vec{F}_N - \frac{1}{2}\vec{F}_N \quad . \tag{19}$$

This means that the pure tensor (spin 2) part of the tensor field gives a repulsive force and only together with the scalar (spin 0) part result is the Newtonian force $\vec{F}_N = \vec{F}_{\{2\}} + \vec{F}_{\{0\}}$. This calculation shows that even on the Newtonian level the physics of the field gravity theory dramatically differs from general relativity.

*The case of static spherically symmetric field.* Substituting Birkhoff's potential (5) with non-linerity correction (15) into the equation of motion (2) one gets the 3-acceleration for a test particle in the frame of the field gravity:

$$\frac{d\vec{v}}{dt} = -(1 + \frac{v^2}{c^2} + 4\frac{\varphi_N}{c^2})\vec{\nabla}\varphi_N + 4\frac{\vec{v}}{c}(\frac{\vec{v}}{c}\bullet\vec{\nabla}\varphi_N) \quad . \tag{20}$$

From the equation of motion (20) it is clear that the acceleration of a test particle depends on the value and direction of its velocity, and this is a coordinate-independent relativistic gravity effect which may be tested experimentally.

*The pericenter shift and positive gravity energy.* The rate of the pericenter shift of the orbit of a test particle with semi-major axis *a*, eccentricity *e* and period *P*, can be directly calculated from the eq.(20):

$$\dot\omega = \frac{6\pi GM}{c^2 a(1-e^2)P} \quad . \tag{21}$$

This formula is the same as in GR, but the interpretation is different. E.g. the nonlinear contribution, i.e. the 2nd term in (15) caused by the positive energy density of the gravity field $T^{00}_{(g)}$, provides 16.7 % of the total value (21).

Therefore in the field gravity theory the pericenter shift is directly affected by the positive energy density of the gravity field, making this physical quantity experimentally measurable.

### Light and atoms interacting with a weak gravity field

*Light in the gravity field.* Within the field gravity theory the deflection of light and the time delay of light signals are consequences of the gravity-electromagnetic field interaction, described by the Lagrangian $L_{(\text{int})} = \psi_{ik}T^{ik}_{(elm)}$. This gives the effective refraction index in the PN approximation:

$$n(r) = 1 + \frac{2GM}{c^2 r} \quad . \tag{22}$$

Hence the velocity of a light signal will have the value

$$c_g(r) = \frac{c}{n} = c(1 - \frac{2GM}{c^2 r}) \quad , \tag{23}$$

So that the direction of light propagation is changed and the time delay appears, both with the same amount as actually observed.

*Atom in gravity field.* The gravitational redshift of spectral lines has another nature than in GR. It is a consequence of the shift of atomic levels. It is universal, because gravitation changes the total energy and all energy levels of an atomic system. In the PN approximation

$$E_{obs} = E_0(1 + \frac{\varphi_N}{c^2}) \quad \text{and} \quad h\nu^{obs}_{ik} = \Delta E^0_{ik}(1 + \frac{\varphi_N}{c^2}) \tag{24}$$

Moshinsky (1950) was the first who calculated the interaction of the gravity field with the spinor and electromagnetic fields of a hydrogen atom. He got the same result as from the energy argument above. A more general formula for the gravitational redshift is:

$$1 + z_g = \frac{1}{\sqrt{1 + \frac{2\phi}{c^2}}} \quad , \tag{25}$$

which gives the correct PN result $z \approx \varphi_N / c^2$.

## 4. Astrophysical tests of the field gravity theory

*Classical relativistic gravity effects*

As we discussed above, from the post-Newtonian approximation of the field gravity theory it follows that classical relativistic gravitational effects - the deflection of light, the gravitational redshift of spectral lines, the time delay of light signals, the perihelion shift, and the Lense-Thirring, Weyl, Schiff precessions, have the same values in GR and FG. Though, the interpretation of the classical effects is different.

This means that one can not make a distinction between geometrical and field approaches just by observing classical relativistic gravity effects in the Solar System and in binary pulsar systems. However, even in the weak field regime there are new, still untested relativistic gravity effects, which may offer crucial experiments for the nature of gravity.

*Testing the equivalence principle*

Modern tests of the equivalence principle achieved the precision in the inferred equality of the inertial and gravitational masses $(m_I = m_G)$ about $10^{-13}$. Several new high-accuracy tests of the equivalence principle have been suggested in last years (Haugan & Lammerzahl 2001; Bertolami, Paramos & Turyshev 2006), which have a goal to discovery a violation of the equivalence principle predicted by modern quantum theories.

Within the field gravity theory the basic concept is the least action principle and the principle of universality of the gravitational interaction (Baryshev 2008a), according to which in the equation of motion (2) the rest mass $m_0$ of a test particle appears in both sides and hence plays the role of inertial and gravitational (passive) mass $(m_0 = m_I = m_G)$. For a body consisting of many particles interacting with each other the most important problem is how to give proper relativistic definitions for inertial and gravitating masses without referring to the non-relativistic Newtonian equation of motion.

According to the PN equation of motion (20) the 3-acceleration of a test particle
- does not depend on the rest mass $m_0$ of the test body,
- depends on the velocity $v$ of the body,
- depends on the value of the gravitational potential $\varphi_N$ at the location of the particle.

This means that there are different ways in relativistic regime to define the inertial $(m_I)$ and the gravitational $(m_G)$ masses. Hence it gives new possibilities to test their equality.

For example a new test of the equivalence principle could utilize the translational motion of a rotating body. According to GR, as a consequence of the equivalence principle, such a body will have the same translational motion as the non-rotating one (if tidal effects can be neglected). However according to FG one should integrate the Poincare gravity force (16) over the volume of the rotating body.

In the case of the translational motion of a rotating body in the weak static spherically symmetric gravitational field the 3-acceleration will be (Baryshev 2002a):

$$\frac{d\vec{V}}{dt} = -(1 + \frac{V^2}{c^2} + 4\frac{\varphi_N}{c^2} + \frac{I\omega^2}{Mc^2})\vec{\nabla}\varphi_N + 4\frac{\vec{V}}{c}(\frac{\vec{V}}{c} \bullet \vec{\nabla}\varphi_N) + \frac{3}{Mc^2}\int [\vec{\omega} \times \vec{r}]([\vec{\omega} \times \vec{r}] \bullet \vec{\nabla}\varphi_N)\,dm \quad . \tag{26}$$

The equation (26) shows that the orbital translational velocity $\vec{V}$ of the center of mass of the body will have extra perturbations due to the rotation. The last term depends on the direction and value of the angular velocity $\vec{\omega}$ of rotation. Its order of magnitude is $v_{rot}^2/c^2$ and it is possible to use this effect for testing the equivalence principle for rotating bodies by astronomical observations with lunar laser ranging (LLR) and timing of pulsars in binary systems.

Baryshev (2002b) calculated the expected violation of the equivalence principle in the case of the LLR experiment as a generalized Nordtvedt effect. Taking into account the rotation of the Earth one can derive that there will be a signal at three frequencies which corresponds to the period D = 29.1 days with amplitude $0.32 \times 10^{-13}$, and at two satellite periods $D_+ = 25.41$ days and $D_- = 35.22$ days with amplitude $0.48 \times 10^{-13}$.

*Deflection from Newtonian law at small masses*

Another test of the nature of the gravitational interaction was suggested by Baryshev & Raikov (1995). Let us compare the gravitational interaction energy $E_{int} = GmM/r$ between two particles (masses $m$ and $M$ at a mutual distance $r$) with the uncertainty principle in the form $\Delta E \, \Delta t > h$. Here the accuracy in measuring the energy is $\Delta E \approx E_{int}$ during the interaction time $\Delta t \approx r/v$, which imply the following condition on the product of masses:

$$mM > \frac{v}{c}\frac{hc}{G} = \frac{v}{c} m_{Pl}^2 \quad . \tag{27}$$

This means that the geodesic motion will be violated if the product of masses is less than the square of the Planck mass $m_{Pl}$ multiplied by $v/c$. So it is expected that for small masses the Newtonian law of the gravity force will be not valid and particle trajectories will have large fluctuations.

If one of the particles is a photon, then it will not be deflected if the wavelength $\lambda$ of the photon is longer than the gravitational radius of the deflecting mass $\lambda > R_g = 2GM/c^2$, so such a photon will not move along a geodesic line. Radio astronomical observations to test this effect were suggested by Baryshev, Gubanov & Raikov (1996).

*Gravitational waves from binary stars*

Gravitational field equation (1) describe the radiation of two types of gravity waves – pure tensor (traceless, spin 2) and scalar (trace of the tensor potential, spin 0). The best test of the validity of the gravitational radiation formulae is offered by binary pulsar systems. For a binary system the loss of energy due to the pure tensor gravitational radiation is given by the quadrupole luminosity (which is the same in field gravity and general relativity):

$$<\dot{E}>_{\{2\}} = \frac{32 G^4 m_1^2 m_2^2 (m_1 + m_2)(1 + \frac{73}{24}e^2 + \frac{37}{96}e^4)}{5 c^5 a^5 (1-e^2)^{7/2}} \quad . \tag{28}$$

Here $m_1$, $m_2$ are masses of the two stars, $a$ is the semimajor axis and $e$ is the eccentricity of the relative orbit.

Within the field gravity theory there is an additional loss of energy due to the scalar monopole radiation (that does not appear in GR), given by the relation (Baryshev 1995):

$$<\dot{E}>_{\{0\}} = \frac{G^4 m_1^2 m_2^2 (m_1 + m_2)(e^2 + \frac{1}{4}e^4)}{4 c^5 a^5 (1-e^2)^{7/2}} \quad . \tag{29}$$

Hence the ratio of the scalar to tensor luminosity is

$$\frac{<\dot{E}>_{\{0\}}}{<\dot{E}>_{\{2\}}} = \frac{5}{128} \frac{(e^2 + \frac{1}{4}e^4)}{(1 + \frac{73}{24}e^2 + \frac{37}{96}e^4)} \quad . \tag{30}$$

The value of this ratio lies in the interval 0 - 1.1 % and for a circular orbit equals zero. However for a pulsating spherically symmetric body there is no quadrupole radiation and the scalar radiation becomes decisive.

According to Damour & Taylor (1991) the orbit of the binary pulsar PSR1913+16 has an eccentricity $e = 0.6171309(6)$, hence the expected scalar radiation contribution (30) is $\Delta_{scalar} = 0.735$ % . Because the rate of change of the orbital period $\dot{P}$ is proportional to the total energy loss, one expects a corresponding excess in the decrease of the orbital period due to scalar gravitational radiation.

The data by Weisberg & Taylor (2002) show that the excess of the orbital period decrease relative to the predicted quadrupole energy loss is $\Delta_{obs} = (observed) - (quadrupole) = 0.78$ % . This is interestingly close to the expected value 0.735 % for the additional energy loss predicted for scalar gravitational radiation (30).

It has been shown by Damour & Taylor (1991) that one must take into account the "Galactic effect" of the accelerations of the pulsar and the Sun in the Galaxy, and that of the proper motion of the pulsar. The distance $d$ to the pulsar PSR1913+16 is a critical parameter in the calculation of the Galactic effect. Unfortunately, the line of sight to the pulsar passes through a complex region of our Galaxy, and one must be very careful, when using known distances to other pulsars for a distance estimate to PSR1913+16.

Damour & Taylor (1991) used indirect arguments to re-estimate the standard dispersion-measure distance of 5.2 kpc. With their new distance d = 8.3 kpc the Galactic effect is +0.69 %, which could almost explain the observed excess. Weisberg & Taylor (2002) took the distance to the pulsar d=5.9 kpc, which gives a Galactic effect of +0.52 %. However, there are also arguments, based on an analysis of the pulse structure of PSR1913+16, lead to a distance of about 3 kpc. For such a short distance the Galactic effect is only +0.11 %.

It is evident that the distance to the pulsar PSR1913+16 requires further investigations. A direct determination of its distance may be regarded as a test of fundamental physics, related to the nature of gravitation. Also distances to other binary pulsars will be crucial for gravity physics.

*Scalar gravitational radiation from supernovae*

*The problem of supernova explosion.* Expected amplitudes and forms of gravitational wave (GW) signals from supernovae explosions detected on the Earth by gravitational antennas essentially depend on the adopted scenario of core-collapsed explosion of massive stars and relativistic gravity theory. This is why the forthcoming GW astronomy will give for the first time experimental limits on possible theoretical models of gravitational collapse including the strong field regime and even quantum nature of the gravity force.

For the estimates of the energy, frequency and duration of supernova GW emission one needs a realistic theory of SN explosion which can explain the observed ejection of massive envelope. Unfortunately, for the most interesting case of SNII explosion such a theory does not exist now. As was noted by Paczynski (1999) if there were no observations of SNII it would be impossible to predict them from the first principles. Modern theories of the core collapse supernova are able to explain all stages of evolution of a massive star before and after the explosion. However, the theory of the explosion itself, which includes the relativistic stage of collapse where a relativistic gravity theory should be applied for the calculation of gravitational radiation, is still controversial and unable to explain the mechanism by which the accretion shock is revitalized into a supernova explosion (see the discussion by Paczynski 1999).

Within the field approach to gravity besides the tensor (spin 2) waves there is the scalar (spin 0) ones, generated by the trace of the energy-momentum tensor of considered matter. For the field gravity theory, there is no detailed calculations of the relativistic stages of the core collapse, but in principle, the energy of scalar GW released by a SN explosion may reach values of about one solar rest mass, with characteristic frequency $10^3$ Hz and durations up to several seconds (Baryshev 1997; Baryshev & Paturel 2001).

*Scalar waves from spherical pulsations of collapsing SN core.* The trace of the tensor field equation (1) gives the field equation for the scalar part $\psi = \eta_{ik}\psi^{ik}$ , which is the usual wave equation:

$$\left(\Delta - \frac{1}{c^2}\frac{\partial^2}{\partial t^2}\right)\psi(\vec{r},t) = -\frac{8\pi G}{c^2} T(\vec{r},t) \quad . \tag{31}$$

The source of the scalar field is the trace of the energy-momentum tensor of collapsing matter in the core of SN. Taking into account the expression for the EMT of the scalar free field, which is

$$T^{ik}_{\{0\}} = \frac{1}{32\pi G}\psi^{,i}\psi^{,k} \quad , \tag{32}$$

and considering the approximation of slow motion in the source, one gets the expression for radiated power in the form of scalar gravitational waves (Baryshev 1997):

$$<\dot{E}>_{\{0\}} = \frac{G}{2c^5}(\dot{E}_{kin})^2 , \tag{33}$$

where the kinetic energy of matter is $E_{kin} = mv^2/2$ .

In particular it follows that in the field gravity theory it is impossible to have a ``quiet'' relativistic collapse of a spherical body because of the violent scalar gravitational radiation. It was shown in Baryshev (1997) and Baryshev & Paturel (2001) that the observed signals from SN 1987A and also detected by Rome bar detectors group (Astone et al. 2002, Pizzella 2008 in this Proceedings), may be understood as a scalar gravitational waves events.

*No black holes in the field approach*

In the case of strong gravity the predictions of FG and GR diverge dramatically. In FG there is no black holes and singularities, and no such limit as the Oppenheimer-Volkoff mass. This means that compact massive objects in binary star systems and active galactic nuclei are good candidates for testing GR and FG theories.

According to FG for a static weak field conditions the positive mass density of the gravitational field around an object with mass $M$ and radius $R$ is

$$\rho_{(g)} = \frac{T^{00}_{(g)}}{c^2} = \frac{(\vec{\nabla}\varphi_N)^2}{8\pi G c^2} = 1.1\times 10^{13}\left(\frac{M}{M_{solar}}\right)^2\left(\frac{10\ km}{R}\right)^4 \ \frac{g}{cm^3} \quad . \tag{34}$$

It is positive, localizable, and does not depend on a choice of the coordinate system. On the surface of a neutron star the mass density of the gravity field is about the same as the mass density of the nuclear matter.

A very general mass-energy argument shows that there cannot be singularities in FG. The total energy of the gravitational field existing around a body is given by

$$E_{(fg)} = \int_R^\infty \frac{1}{8\pi G}(\vec{\nabla}\varphi_N)^2 4\pi r^2 dr = \frac{GM^2}{2R} \quad . \tag{35}$$

This energy should be less than the rest mass energy of the body, which includes the energy of the gravity field. From this condition it follows that:

$$E_{(fg)} < Mc^2 \quad \Rightarrow \quad R > \frac{GM}{2c^2} \quad . \tag{36}$$

If one takes into account the non-linearity of the gravity field, then the value of the limiting radius further increases, because "the energy of the field energy" should be added. Hence a safe estimate for the limiting minimum radius of any massive body in the field gravity is $R_m > 0.5 \, r_g$. This argument is a precise analogue to that of the classical radius of electron $R_e = e^2 / m_e c^2$, following from the requirement that the electric field energy $E_{fe} = e^2 / 2R_0$ should be less than the electron's rest-mass energy $E_0 = m_e c^2$.

Thus black holes and singularities are excluded by the existence of the positive energy density of the gravitational field.

*The limit on the gravity force*

The positive energy-density of the gravitational field leads to a limit on the gravity force acting on a test body from an object having the limiting radius $R_M = GM/c^2$. Indeed, in the weak field approximation the field equation outside a body with mass $M$, surrounded by a positive field energy density (34), should take into account the source term caused by this mass-energy:

$$\Delta \varphi = +\frac{1}{c^2}(\vec{\nabla}\varphi)^2 , \quad \text{and its solution} \quad \varphi = -c^2 \ln(1 + \frac{GM}{c^2 r}) \ . \tag{37}$$

Hence the gravity force will be

$$F_g = m\frac{d\varphi}{dr} = \frac{GmM}{r^2}\frac{1}{(1+\frac{GM}{c^2 r})} \ . \tag{38}$$

For a maximally compact relativistic object having the radius $R_M = GM/c^2$, the gravitational acceleration and the gravity force are restricted by

$$g_{max} \leq \frac{c^4}{GM} = \frac{c^2}{R_M} \quad \text{and} \quad F_g \leq \frac{mc^4}{GM} = \frac{mc^2}{R_M} < \frac{c^4}{G} \ , \tag{39}$$

where the last equality is written for the case $m = M$.

In general relativity the energy-density of the gravity field is negative (see discussion in preceding paper), hence the sign of the right-hand side of the field equation is negative and in this case

$$\Delta \varphi = -\frac{1}{c^2}(\vec{\nabla}\varphi)^2 , \quad \text{and its solution} \quad \varphi = -c^2 \ln(1 - \frac{GM}{c^2 r}) \ . \tag{37}$$

So the gravity force will be

$$F_g = m\frac{d\varphi}{dr} = \frac{GmM}{r^2}\frac{1}{(1-\frac{GM}{c^2 r})} \ , \tag{38}$$

And for $r \to R_M$ the gravity force is infinite at finite radius. This difference in the behavior of the gravity force in GR and FG has important consequences for the structure and stubility of relativistic astrophysical objects.

*Hydrostatic equilibrium configurations*

In general relativity the Tolman-Oppenheimer-Volkoff equation of hydrostatic equilibrium leads to a maximum mass of a neutron star, about 2 solar masses, called the Oppenheimer-Volkoff limit. Larger masses can exist only in the form of black holes.

In the field gravity theory the equations of motion are contained in the conservation laws $T^{ik}_{,k} = 0$, where $T^{ik} = T^{ik}_{(p/m)} + T^{ik}_{(int)} + T^{ik}_{(g)}$, is the total EMT of considered system gas + gravity field in corresponding approximation. The post-Newtonian equation of hydrostatic equilibrium in FG was derived by Baryshev (1988). It depends on a particular choice of the interaction EMT and may be written in the form

$$\frac{dp}{dr} = -\frac{G(\rho_0 + \delta\rho)(M_0^r + \delta M^r)}{r^2} \tag{39}$$

where $\delta\rho = (e+p)/c^2 + 2\rho_0\phi$, and $\rho_0$ is the rest-mass density, $\phi = \psi^{00}$, $M_0^r = \int_0^r 4\pi r^2 dr$,

$\delta M^r = \int_0^r 4\pi r^2 [(e+3p)/c^2 + 2\rho_0\phi/c^2 + (d\phi/dr)^2/8\pi G]\, dr$. The most important difference between equations of hydrostatic equilibrium in FG and in GR is that within FG the relativistic gravity corrections lead to a decrease of the gravitating mass relative to the rest-mass due to the negative value of the gravitational potential ($\phi < 0$). According to eq.(39) a hydrostatic equilibrium is possible for any large mass.

The internal structure of the neutron stars within FG was numerically studied by Tanychin (1995), who showed that in FG the stars are more homogeneous than in GR, and that there is no upper limit on their masses.

*Stability of supermassive stars in field gravity theory*

Hoyle & Fowler (1963) suggested that a mass of the order of $10^8\, M_S$ (solar masses) may condense in a galactic nucleus into a supermassive star (SMS), in which the nuclear energy generation takes place. However, a year later Fowler (1964) showed that in general relativity a SMS is unstable and will collapse to a black hole within a lifetime $\tau \approx 10(M/10^8 M_S)^{-1}$ yr before the nuclear reactions begin. Hence in the standard GR only black holes can be the primary power sources of the active galactic nucleus.

Within the field gravity theory the SMS is stable, which was shown by Baryshev (1992) using the method developed by Fowler (1966) for considering the PN hydrostatic equilibrium and small adiabatic pulsations of a slowly rotating SMS. The total equilibrium energy of SMS (excluding the constant term $\int \rho_0 c^2 dV = M_0 c^2$ has the form

$$E^{(equi)} = \int_V (e - 3p - \frac{1}{2}\rho_0 v^2)\, dV, \tag{40}$$

which is a consequence of the relativistic virial theorem in the PN approximation. Here $e$ is the thermal energy density, $p$ is the pressure, $(1/2)\rho_0 v^2$ is the density of the kinetic rotational energy, so that $\Delta E_{rot} = \int (1/2)\rho_0 v^2 dV$. The first two terms in (40) can be expressed via the Newtonian potential energy $E_{pot}$ plus the relativistic correction $\Delta E_{rel}$, hence

$$E^{equi} = \frac{\beta}{2} E_{pot} - \Delta E_{rel} - \Delta E_{rot}, \tag{41}$$

where $\beta = p_{gas}/p_{tot} \ll 1$ is the gas to the total pressure ratio ($\beta \approx 10^{-3}$ for a SMS mass $10^8\, M_S$) and the relativistic correction is

$$\Delta E_{rel} = K(n, gt)\frac{G^2}{c^2} M_0^{7/3} \rho_{0c}^{2/3}, \tag{42}$$

where $M_0$ is the rest-mass of the SMS, $\rho_{0c}$ is the central mass density, $K(n, gt)$ is a constant defined by the polytrope index $n$ and the gravity theory.

The stability of the SMS follows from the fact that in FG the constant $K>0$, while in GR the constant $K<0$. For the $n = 3$ polytrope the calculation of the value of $K(3, FG)$ within the field gravity theory was done by A.Raikov (details in Oschepkov \& Raikov 1995): $K(3, FG) = + 1.7349$, while in GR we have $K(3, GR) = - 0.9183$. The different signs of the relativistic gravity corrections for GR and FG show that in general relativity we have a PN instability, while in field gravity the supermassive star is stable. The PN stability of the SMS in the field gravity theory radically changes the understanding of their evolution. In particular, at the last stages of the SMS evolution the main energy source will not be nuclear reac-tions which produce the energy output of about 1 % of $M_0 c^2$, but the gravitational binding energy of the order of total $M_0 c^2$.

**5. Conclusion: crucial astrophysical tests of gravity physics**

*Relativistic compact objects*

Observations of relativistic compact objects (RCO), also called as "black hole candidates", in the X-ray binary stellar systems (RCO masses about 10 solar masses) and in galactic nuclei ($10^6 - 10^9$ solar masses) provide astrophysical tests of the strong gravity effects and hence the nature of the gravitational interaction. Within the geometrical approach, classical general relativity predicts black holes --- theoretical objects having the event horizon at the Schwarzchild radius $r = R_g = 2GM/c^2$, after which a one-way fall into the singularity is inevitable. This radically new physics has no counterpart in all other fundamental physical interactions.

There is no singularity and event horizon in the frame of the field gravity theory. As we discussed above, the positive localizable energy density of the gravity field prevents the appearance of a singularity at the center and also at the gravitational radius of a RCO. This is strictly the same physical reason as the absence of the singularity of an electron in electrodynamics. Instead of black holes, the field gravity theory predicts the existence of massive compact relativistic objects having radii $r = R_m = GM/c^2$, close to the gravitational radius. Field gravity RCOs have a highly redshifted surface and an intrinsic magnetic field.

To prove or disprove observationally the existence of black holes means to prove or disprove the existence of the event horizon in relativistic compact objects. Crucial observational tests, which would convincingly show the existence of the event horizon around a RCO, have not yet been made.

A discussion about the "visibility" of the event horizon of black holes is going on in the literature. It is difficult to prove the existence of such a one-way sphere, because of many astrophysical processes are involved. It has been even stated (Abramowicz et al. 2002), that it is impossible to prove observationally that an object has an event horizon. The most difficult for black hole models is to explain the observed very small luminosity in a certain variability phase, when the accretion rate is still large enough.

Narayan & Quataert (2005) suggested that the low luminosity could be explained by introducing a new physical process, the "advection dominated accretion flow (ADAF)" or "radiatively inefficient accretion flow (RIAF)". ADAF is based on the assumption that protons and electrons are decoupled in the flow, so the kinetic energy is absorbed by the event horizon without an outward radiation. However, it was noted by Bisnovatyi-Kogan \& Lovelace (2000) and Binney (2003) that the magnetic field present in astrophysical plasmas of the accretion flow make ADAF practically impossible.

Robertson & Leiter (2002, 2003) analyzed observational data on black hole candidates in X-ray binary stars and active galactic nuclei and found evidence for intrinsic magnetic fields, which is in conflict with the black hole model. The low luminosity phase is naturally explained by the "propeller effect" of the magnetic field of the RCO. In the frame of general relativity, Robertson and Leitner used a new RCO model, the "magnetospheric eternally collapsing object (MECO)" that has no event horizon though its size is close to the Schwarzschild radius (see Mitra 2008 in this Proceedings). Observations of the gravitationally lensed quasar Q0957+561A,B revealed the inner structure of the accretion disc, which demands an intrinsic magnetic field of the central RCO and may be well modeled by the MECO (Schild, Leiter & Robertson 2006, Schild 2008 in this Proceedings). Note, that the field gravity RCO also can explain the intrinsic magnetic fields in the galactic "black hole candidates" and active galactic nuclea.

Another unexpected finding in the RCO observations is the very small radius of radiating matter in accretion discs. E.g., in the best studied accretion disc, around the central object in the Sy1 galaxy MCG-6-30-15, the inner radius of the orbiting matter is $r_{inner} = 0.615 \, R_g$. This is less than the Schwarzschild radius and within general relativity it had to be interpreted as an extremely rotating Kerr black hole (Wilms et al. 2001).

Crucial observational tests, capable of distinguishing between the alternative models for RCO, are difficult. Perhaps the most direct test of the black hole model was suggested by Falcke, Melia \& Agol (2000), who discussed VLBI observations of the black hole candidate in the Galactic center with micro-arcsecond angular resolution. The profile of such an image can even distinguish between non-rotating and rotating black holes.

*Core-collapse supernovae, gamma-ray bursts and gravitational waves*

Another direct test of the strong gravity effects would be the detection of a gravity wave signal from the relativistic collapse. The absence of black holes in the field gravity makes dramatic changes in the physics of supernova explosions. The collapse of the iron core of massive pre-supernovae stars will have a pulsation character and leads to long duration gravitational signals, comparable with neutrino signals and gamma ray bursts, i.e. several seconds.

The relation of the gamma-ray-burst (GRB) phenomenon to relativistic core-collapse supernovae has become a generally accepted interpretation of the GRBs (Paczynski 1999, Sokolov 2008 in this Proceedings). If the compact GRB model suggested by Sokolov et al. (2006) obtains further confirmation, then there should be a correlation of the gamma-x-ray signal with gravitational bursts. The gravitational antenna GEOGRAV observed a signal from SN1987A (Amaldi et al. 1987) together with the neutrino signal observed by the Mont Blanc Underground Neutrino Observatory (Aglietta et al. 1987, Pizzella 2008 in this Proceedings). This has been interpreted by Baryshev (1997) as a possible detection of the scalar gravitational radiation from the spherical core-collapse of the pre-supernova. An observational strategy to distinct between scalar and tensor gravitational waves by using siderial time analysis was considered by Baryshev & Paturel (2001) and Paturel & Baryshev (2003a,b). Recent, still controversial, claims about possible detections of gravitational signals by Nautilus and Explorer antennas (Astone et al. 2002, Pizzella 2008 in this Proceedings), if confirmed, require a new analysis of the potential sources of gravitational waves (Coccia, Dubath & Maggiore 2004).

*Applications to cosmology*

Observational cosmology provides the possibility to study matter distribution and its evolution on largest achievable scales. Such observations also test gravity theories in their ability to describe the whole Universe. The geometrical approach of general relativity leads to the Friedmann cosmological model, the standard frame for modern cosmological research. The expanding homogeneous universe explains all available data, though suffering from some paradoxes, which is discussed in this Proceedings.

The field gravity theory allows one to operate with a matter distribution in infinite Minkowski space without the gravitational potential paradox. A global evolution of matter is possible without space expansion and initial singularity. Cosmological redshift could have gravitational nature. The energy-momentum tensor of the interaction plays the role of an effective cosmological $\Lambda$ term (Baryshev 2008c in this Proceedings ).